\documentclass[a4paper,fleqn,10pt]{article}
\usepackage[numbers,sort&compress]{natbib}
\usepackage{graphicx}
\usepackage{subfigure}
\usepackage{latexsym}
\usepackage{amsmath,amssymb}
\usepackage{dcolumn}
\usepackage{float}
\usepackage{caption}

\begin{document}
\captionsetup[figure]{labelfont={bf},labelformat={default},labelsep=space,name={Fig.}}

\title{{\bf Orbital motion of test particles in regular Hayward black hole space-time}
\author{\normalsize Jian-Ping Hu$^{1}$ and Yu Zhang$^{1}$\thanks{Corresponding Author(Y. Zhang): Email: zhangyu\_128@126.com}\\
   \normalsize \emph{$^{1}${Faculty of Science, Kunming University of Science and Technology, }}\\ \normalsize \emph {Kunming, Yunnan 650500, People's Republic of China}}}

\date{}
\maketitle \baselineskip 0.3in

{\bf Abstract} \ {In this paper, all possible orbits of test particles are investigated by using phase plane method in regular Hayward black hole space-time. Our results show that the time-like orbits are divided into four types: unstable circular orbits, separates stable orbits, stable hyperbolic orbits and elliptical orbits in regular Hayward black hole space-time. We find that the orbital properties vary with the change of $\ell$ (a convenient encoding of the central energy density $3/8\pi\ell^{2}$). If $\ell =\frac{1}{3}$ and $b < 3.45321$, the test particles which moving toward the black hole will definitely be plunging into the black hole. In addition, it is obtained that the innermost stable circular orbit happens at $r_{min}$ = 5.93055 for $b$ = 3.45321.}



\section{Introduction}
In 1968, Bardeen proposed the first regular black hole model named Bardeen black hole\cite{bardeen1968non} to
avoid singularity problem\cite{virbhadra2007time,frolov2008singularity,khodadi2016more}. Since then, many other regular models are presented, for example the regular Hayward black hole\cite{hayward2006formation,abbas2014geodesic}. In this black hole model, the space-time metric is built up on the Schwarzschild metric\cite{virbhadra2000schwarzschild,cardoso2003quasinormal,konoplya2005scalar,Ibrar2016Marginally} and a new parameter $\ell$ is introduced. When $\ell = 0$, this metric reverts to the metric of Schwarzschild black hole space-time. The type of horizon is determined by the values of ${\frac{\ell^{2}}{m^{2}}}$($m$ is the mass of regular Hayward black hole). When ${\frac{\ell^{2}}{m^{2}}}>\frac{16}{27}$, no horizon (regular particle solution) will be formed. When ${\frac{\ell^{2}}{m^{2}}} = \frac{16}{27}$, single horizon (regular extremal black hole)solution is allowed. When ${\frac{\ell^{2}}{m^{2}}} < \frac{16}{27}$, double horizons (regular black hole with two horizons) are allowed\cite{halilsoy2014thin}. The variations of parameter $\ell$ can affect the properties of gravitational field, but the weak energy condition\cite{bardeen1973four,ayon1998regular} is always satisfied, just as in Bardeen black holes. Researches on the regular Hayward black hole\cite{chiba2017anote,abdujabbarov2017gravitational} are the focus of black hole physics all the time. Some researchers\cite{lin2013quasinormal,flachi2013quasinormal,toshmatov2015quasinormal} devoted to study of the quasinormal modes (QNMs) of the regular Hayward black hole. Halilsoy et al.\cite{halilsoy2014thin} found that a Hayward parameter $\ell$ could make the Thin-shell wormhole\cite{sharif2014stability} more stable. In 2014, Abbas and Sabiullah\cite{abbas2014geodesic} demonstrated the timelike and null geodesic structures\cite{cardoso2009geodesic,chen2009timelike,muller2011studying,farrugia2017thermodynamic,azam2017geodesic,azam2017geodesic2} of massless (photon) and massive particles in regular Hayward space-time. Soon, Amir and Ghosha\cite{amir2015rotating} suggested that a rotating regular Hayward black hole could also act as a particle accelerator. Recently, Amir et al.\cite{amir2016collision} presented that the ergoregion will be enlarged when the values of $g$ (the deviation parameter) increases in regular Hayward black hole space-time.

In this paper, we make some analyses for the orbital motions of test particles\cite{hussain2015a} in regular Hayward black hole space-time. We get a very clear picture to describe different types of the orbital motion by using the phase plane method\cite{dean1999phase}. Our results can help us better understand the geometrical properties of the regular Hayward black hole space-time.

The structure of the paper is as follows: In Sect.2, we derive the equations of the orbital motion. In Sect.3, the stability\cite{rosa2007stability} of the orbital motion is studied using the phase plane method. In Sect.4, the effect of the parameter $\ell$ on orbital motion is investigated. In Sect.5, a brief conclusion is given.

\section{Equations of the orbital motion}
The metric of regular Hayward black hole space-time is described by\cite{hayward2006formation}
\begin{eqnarray}
ds^{2}=-f(r)dt^{2}+\frac{1}{f(r)}dr^{2}+r^{2}d\theta^{2}+r^{2}\sin^{2}\theta d\phi^{2},
\label{e1}
\end{eqnarray}
\begin{eqnarray}
f(r)=1-\frac{2mr^{2}}{r^{3}+2ml^{2}},
\label{e2}
\end{eqnarray}
where $m$ is the mass of regular Hayward black hole and $\ell$ is a convenient encoding of the central energy density $3/8\pi\ell^{2}$\cite{abbas2014geodesic,halilsoy2014thin,lin2013quasinormal,amir2015rotating}. We choose $m = 1$, $\ell = \frac{1}{3}$ to get the double horizons (regular black hole with two horizons).
The Lagrangian for test particles can be written as
\begin{eqnarray} L=\frac{1}{2}M(\frac{ds}{d\tau})^2=-\frac{1}{2}Mf(r)\dot{t}^2+\frac{1}{2}M\frac{1}{f(r)}\dot{r}^2+\frac{1}{2}Mr^2\dot{\theta}^2+\frac{1}{2}Mr^{2}\sin^{2}\theta\dot{\phi}^2,
 \label{e3}
\end{eqnarray}
where $\tau$ is the proper time, $M$ is mass of the test particles. $\dot{t} = dt/d\tau$,\  $\dot{r} = dr/d\tau$, \ $\dot{\theta} = d\theta/d\tau$, \ $\dot{\phi} = d\phi/d\tau$.

Without losing generality, we choose $\theta = \frac{\pi}{2}$. Eq. (\ref{e3}) can be expressed as
\begin{eqnarray}
L=-\frac{1}{2}Mf(r)\dot{t}^2+\frac{1}{2}M\frac{1}{f(r)}\dot{r}^2+\frac{1}{2}Mr^{2}\dot{\phi}^2.
 \label{e4}
\end{eqnarray}
\begin{eqnarray}
\frac{d}{d\tau}(\frac{\partial L}{\partial \dot{x}^{\nu}})-\frac{\partial L}{\partial x^{\nu}}=0.
\label{e5}
\end{eqnarray}
Considering the equation of Lagrangian not explicitly containing $t$ and $\phi$, by using the Euler-Lagrangian differential equation\cite{muslih2005hamiltonian} (Eq. (\ref{e5})), we can obtain two equations
\begin{eqnarray}
\frac{\partial L}{\partial t}=0 \Longrightarrow{-\frac{\partial L}{\partial \dot{t}}=\varepsilon=Mf(r)\dot{t}},
\label{e6}
\end{eqnarray}
\begin{eqnarray}
\frac{\partial L}{\partial \phi}=0 \Longrightarrow{-\frac{\partial L}{\partial \dot{\phi}}=J=Mr^{2}\dot{\phi}}.
\label{e7}
\end{eqnarray}
By analyzing these two equations, two constants $\varepsilon$ (the total energy) and $J$ (the total angular momentum) of the test particles are obtained.\\
Let $E = \frac{\varepsilon}{M}, \ b = \frac{J}{M}$ to simplify Eqs. (\ref{e6}) and (\ref{e7})
\begin{eqnarray}
\dot{t}=\frac{E}{f(r)}, \ \ \ \dot{\phi}=\frac{b}{r^{2}},
\label{e8}
\end{eqnarray}
then we can get
\begin{eqnarray}
\dot{r}^{2}=E^{2}-\frac{b^{2}}{r^{2}}f(r)-f(r).
\label{e9}
\end{eqnarray}
Due to  $\dot{r} = \frac{dr}{d\phi}\dot{\phi}$ and  $\dot{\phi} = \frac{b}{r^{2}}$, Eq. (\ref{e9}) can be rewritten as
\begin{eqnarray}
(\frac{dr}{d\phi})^{2}\frac{b^{2}}{r^{4}}=E^{2}-\frac{b^{2}}{r^{2}}f(r)-f(r).
\label{e10}
\end{eqnarray}
 Define $R = \frac{r_{+}}{r}$ (here $\ell = \frac{1}{3}$), we have
 \begin{eqnarray}
 (\frac{dR}{d\phi})^{2}=\frac{r_{+}^{2}}{b^{2}}(E^{2}-1)-R^{2}+R^{2}g(R)+\frac{r_{+}^{2}}{{b}^{2}}g(R),
\label{e11}
\end{eqnarray}
where $g(R)=\frac{2mRr_{+}^{2}}{r_{+}^{3}+\frac{2}{9}R^{3}m}$.
\section{Stability of the orbital motion}
Now, we would like to make a research about the properties of the orbital motion in regular Hayward black hole
space-time by analyzing the effective potential\cite{chen2009timelike,fernando2012schwarzschild,sheng2011time-like} and using the phase plane method. First of all, define $x = R\ and\ y = \frac{dR}{d\phi}$, then Eq. (\ref{e11}) would become
\begin{eqnarray}
y^{2}=\frac{r_{+}^{2}}{b^{2}}(E^{2}-1)-x^{2}+x^{2}g(x)+\frac{r_{+}^{2}}{{b}^{2}}g(x),
\label{e12}
\end{eqnarray}
where $g(x)=\frac{2mxr_{+}^{2}}{r_{+}^{3}+\frac{2}{9}x^{3}m}$. When $\frac{dR}{d\phi}=0$, the effective potential can be defined as
\begin{eqnarray} V_{eff}^{2}-1=\frac{b^{2}x^{2}}{r_{+}^{2}}(1-\frac{2mxr_{+}^{2}}{r_{+}^{3}+\frac{2}{9}x^{3}m})-\frac{2mxr_{+}^{2}}{r_{+}^{3}+\frac{2}{9}x^{3}m}.
\label{e13}
\end{eqnarray}
Up to now, we have obtained the equation of the orbital motion and the equation of the effective potential. When the effective potential satisfy the condition $\frac{dV_{eff}^{2}}{dx} = 0$, the orbits would be circular. The form of the
condition in regular Hayward black hole is as follows
\begin{eqnarray}
 \frac{dV_{eff}^{2}}{dx}=\frac{4m^{2}r_{+}^{2}x^{3}}{3(r_{+}^3+\frac{2mx^3}{9})^{2}}-\frac{2mr_{+}^{2}}{r_{+}^3+\frac{2mx^{3}}{9}}+\frac{b^{2}x^{2}(\frac{4m^{2}r_{+}^{2}x^{3}}{3(r_{+}^{3}+\frac{2mx^{3}}{9})^{2}}-\frac{2mr_{+}^{2}}{r_{+}^3+\frac{2mx^{3}}{9}})}{r_{+}^{2}}+\frac{2b^{2}x(1-\frac{2mxr_{+}^{2}}{r_{+}^3+\frac{2mx^{3}}{9}})}{r_{+}^2}=0.
 \label{e14}
 \end{eqnarray}

 And the stability condition of the test particles requires

 \begin{equation}
   \begin{split}
 \frac{d^{2}V_{eff}^{2}}{dx^{2}}=&-\frac{16m^{3}r_{+}^{2}x^{5}}{9(r_{+}^{3}+\frac{2mx^{3}}{9})^{3}}+\frac{16m^{2}r_{+}^{2}x^{2}}{3(r_{+}^{3}+\frac{2mx^{3}}{9})^{2}}+\frac{b^{2}x^{2}(-\frac{16m^{3}r_{+}^{2}x^{5}}{9(r_{+}^{3}+\frac{2mx^{3}}{9})^{3}}+\frac{16m^{2}r_{+}^{2}x^{2}}{3(r_{+}^{3}+\frac{2mx^{3}}{9})^{2}})}{r_{+}^{2}}\\
 &+\frac{4b^{2}x(\frac{4m^{2}r_{+}^{2}x^{3}}{3(r_{+}^{3}+\frac{2mx^{3}}{9})^{2}}-\frac{2mr_{+}^{2}}{r_{+}^{3}+\frac{2mx^{3}}{9}})}{r_{+}^{2}}+\frac{2b^{2}(1-\frac{2mr_{+}^{2}x}{r_{+}^{3}+\frac{2mx^{3}}{9}})}{r_{+}^2} \geq 0.
 \label{e15}
 \end{split}
 \end{equation}
  \begin{figure}[H]
\centering
    \includegraphics[angle=0, width=0.4\textwidth]{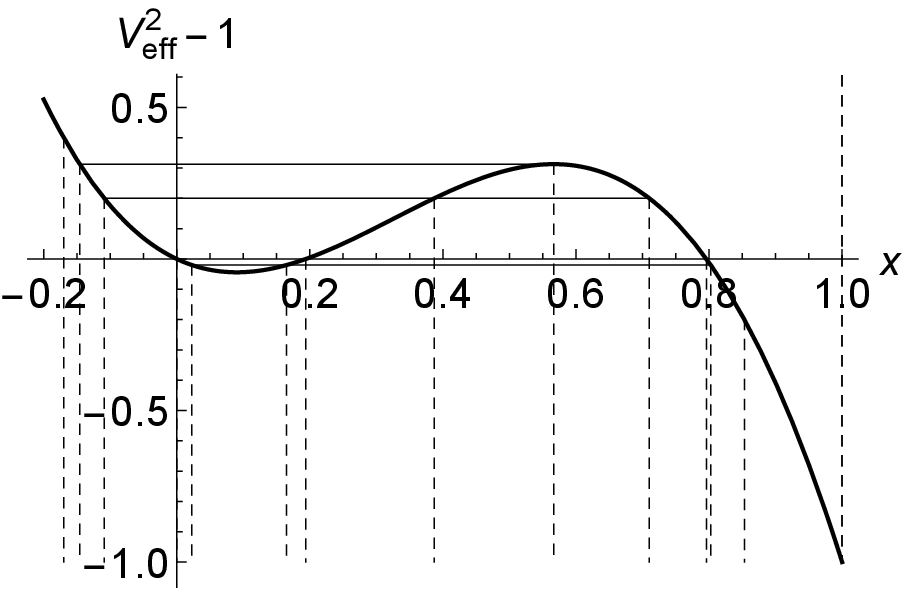}

    \ \includegraphics[angle=0, width=0.4\textwidth]{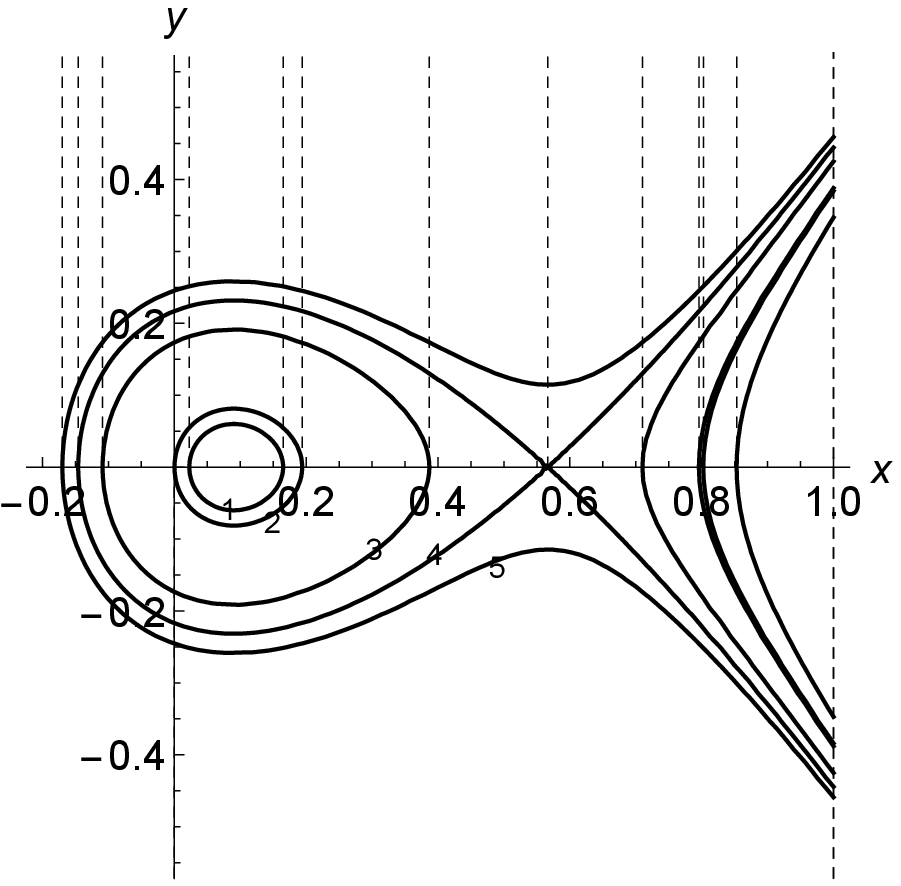}
     \vspace*{8pt}
\caption{The behaviors of the effective potential and the phase plane of particles with $b = 5$, $m = 1$ and $\ell = \frac{1}{3}$.  \label{v1} }
\end{figure}
In Fig. \ref{v1} we plot the effective potential graph and the phase plane graph with the angular momentum $b = 5$. According to Eq. (\ref{e12}), we choose the values of energy $E^{2}-1$ = -0.02, 0, 0.2, 0.31281, 0.4 corresponding to orbits 1 to 5, respectively. In terms of $\frac{dV_{eff}^{2}}{dx} = 0$, we get the orbit 4 (unstable circular orbit). It is a dividing curve between the stable orbits and the unstable orbits. Orbits outside it are unstable, such as orbit 5. Orbits inside it are stable, such as orbit 1. For $E^{2}-1=0$, we find that it corresponds to orbit 2 which separates stable orbits into two categories: orbits outside it are the stable hyperbolic type curves(for example orbit 3 ) and inside it are the elliptical type curves(for example orbit 1).

\begin{figure}[H]
\centering
    \includegraphics[angle=0, width=0.4\textwidth]{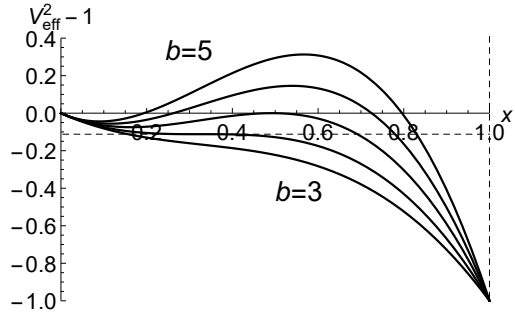}
    \vspace*{8pt}
\caption{The effective potential changing with the angular momentum. From bottom to top, $b = 3$, $3.45321$, $3.98578$, $4.5$, $5$. We choose $m = 1$,$\ell = \frac{1}{3}$. \label{v2} }
\end{figure}

\begin{figure}[H]
 \centering
    \includegraphics[angle=0, width=0.25\textwidth]{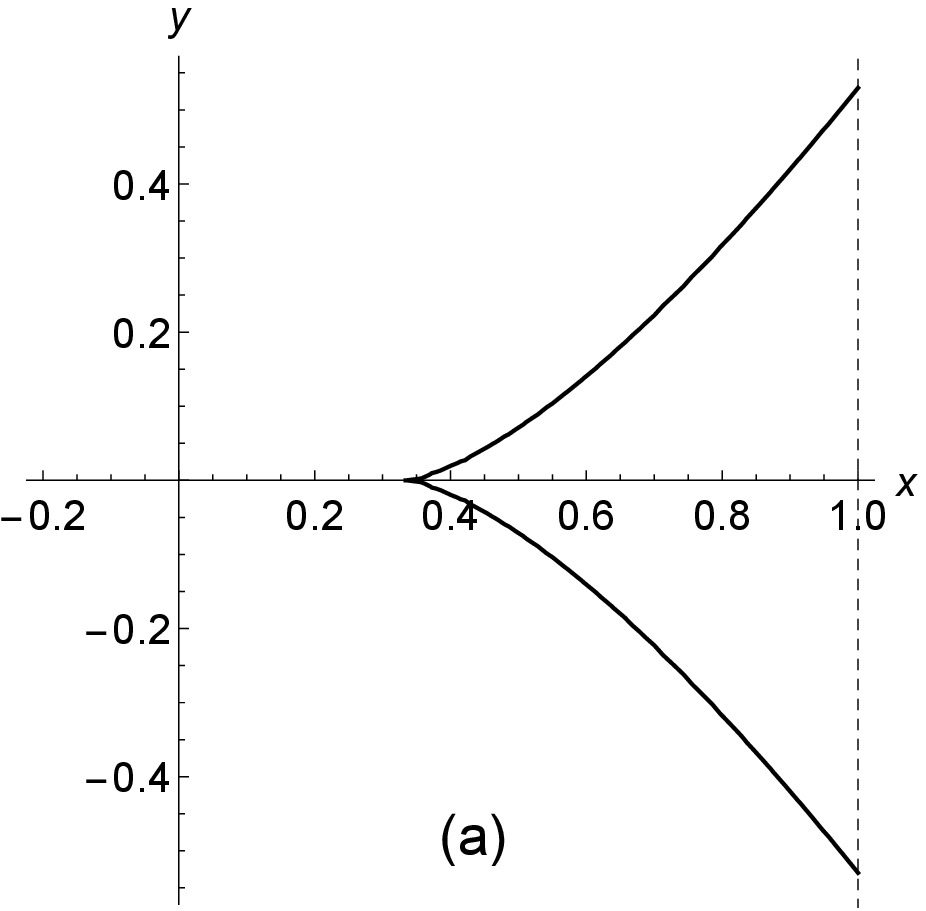}
        \includegraphics[angle=0, width=0.25\textwidth]{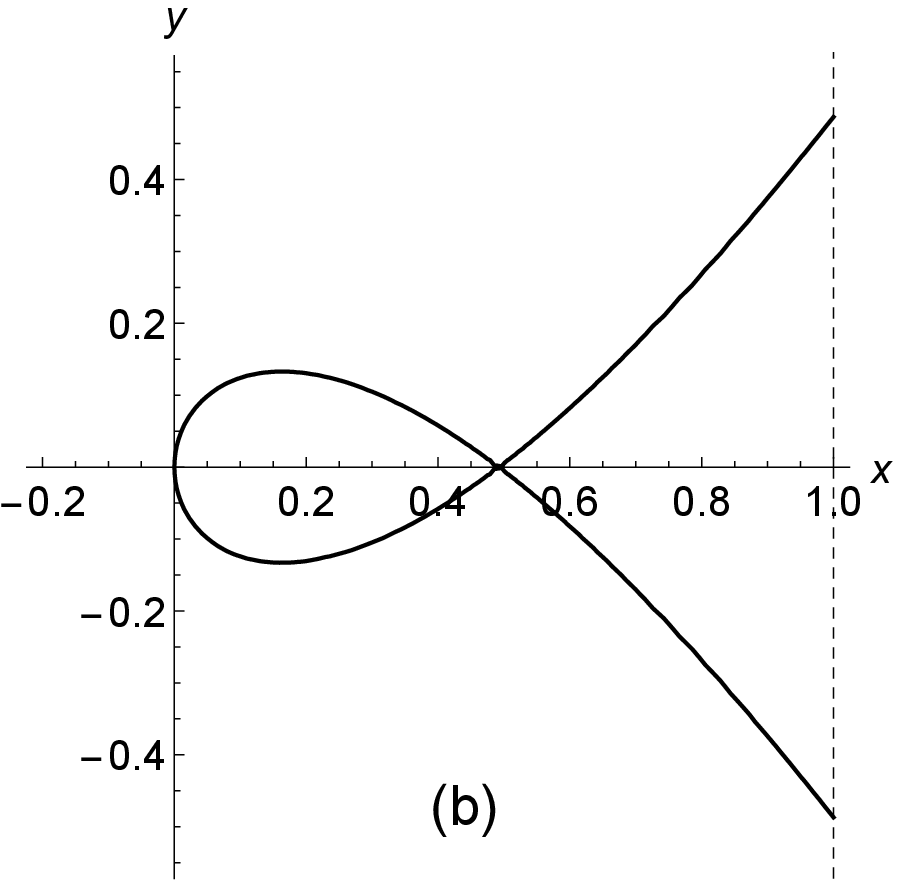}\\
        \includegraphics[angle=0, width=0.25\textwidth]{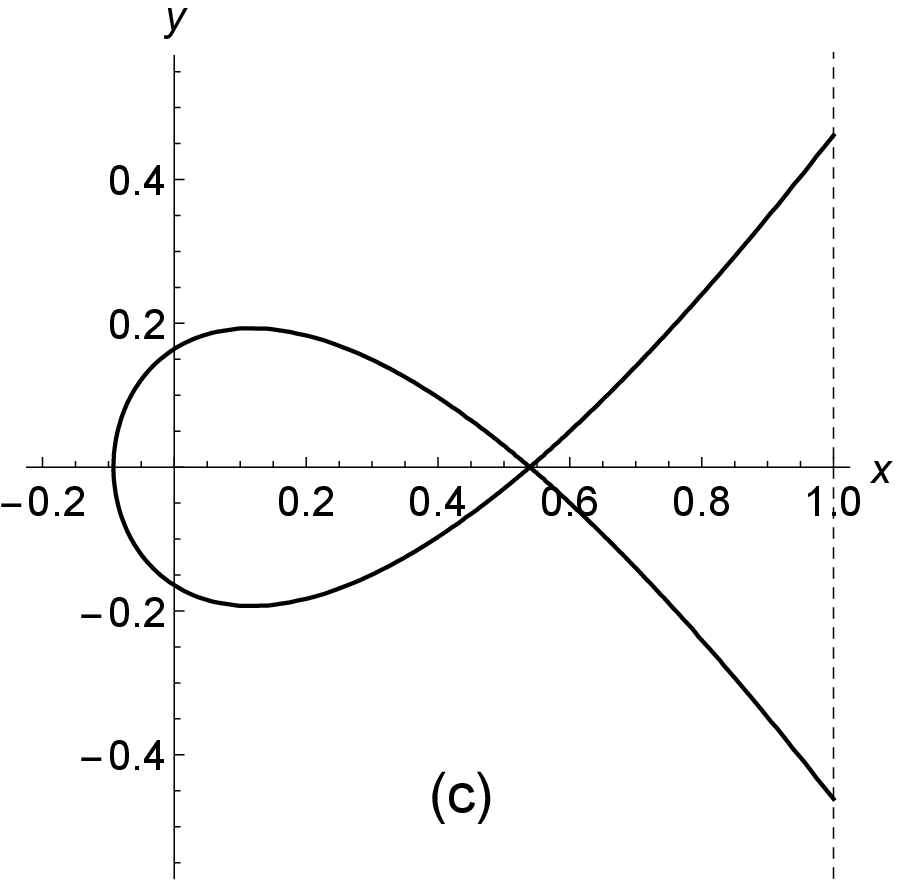}
        \includegraphics[angle=0, width=0.25\textwidth]{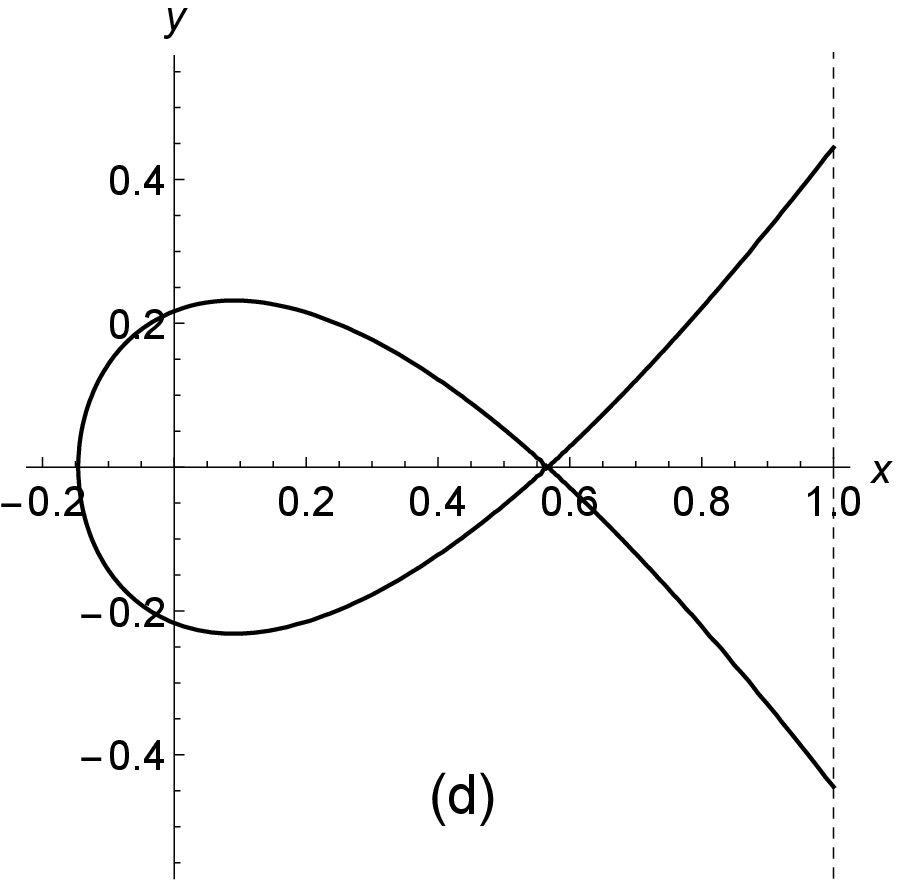}
        \vspace*{8pt}
\caption{Dividing line of orbital stability with different values of $b$ :\ (a) $b$ = 3.45321, (b) $b$ = 3.98578, (c) $b$ = 4.5, (d) $b$ = 5. We choose $m = 1$\ and\ $\ell = \frac{1}{3}$. \label{v3} }
\end{figure}

The motion of test particles is not only influenced by particle energy but also particle angular momentum. In Fig. \ref{v2}, we present the dependence of the effective potential on angular momentum. Here we choose $b = 3$, $3.45321$, $3.98578$, $4.5$, $5$. The special value $b = 3.45321$ is obtained by analyzing two equations $\frac{dV_{eff}^{2}}{dx} = 0$ and $\frac{d^{2}V_{eff}^{2}}{dx^{2}}$ = 0. When $b < 3.45321$, the curve of the effective potential has no extremal point. In this case, it is not possible to have a stable orbit. No matter we throw the test particles which have how much energy in the black hole direction, they will always plunge into the black hole finally. We find that the values of the angular momentum determines the types of orbits.
In order to contrast orbital properties about different values of $b$, we choose $b$ = $3.45321$, $3.98578$, $4.5$, $5$ corresponding to $V^{2}_{eff}-1$ = $-0.11205$, $0$, $0.14564$, $0.31281$ to plot Fig. \ref{v3}. By comparing the four images in Fig. \ref{v3}, we get different radius of the innermost stable circular orbit. Combining the above three graphs to analyze, we draw the following conclusions:

\begin{enumerate}
  \item When $b < 3.45321$, from Fig. \ref{v2} we see that the curve of the effective potential\ ($V^{2}_{eff}-1$) has no extremal point, which means there is no stable orbit. It is not difficult to find that the angular momentum of test particles is not large enough to maintain a stable motion. When we throw test particles in black hole direction, the only result is to be absorbed by black holes.
  \item When $b$ = 3.45321, the curve of the effective potential($V^{2}_{eff}-1$) has one inflection point which corresponds to the minimum stable circular orbit. The radius of the innermost stable circular orbit is $r_{min} = 5.93055.$

  \item When $ 3.45321 < b < 3.98578$, the maximum value of $V^{2}_{eff}-1$ is negative. According to Figs. \ref{v1} and \ref{v3}, both stable and unstable orbits exist. In terms of the distinguish condition $E^{2}-1 = 0$, stable elliptical orbit is the only choice.

  \item When $b = 3.98578$, the maximum value of $V^{2}_{eff}-1$ is 0. There exist stable and unstable orbits. According to the distinguish condition $E^{2}-1 = 0$, we can draw a conclusion that there is one stable hyperbolic orbit and others are elliptical.

  \item When $b > 3.98578$, the maximum values of $V^{2}_{eff}-1$ is positive. There are stable and unstable orbits. And the stable orbits have two types including elliptical orbit and hyperbolic orbit.
\end{enumerate}

\section { The influence of the parameter $\ell$ }
From Eqs. (\ref{e1}) and (\ref{e2}), we find that the regular Hayward black hole space-time varies when the values of parameter $\ell$ is changed. We use the same test particles which have equal values of $b$ to investigate the influence of parameter $\ell$ on the orbital motions.

\begin{figure}[!ht]
\centering
    \includegraphics[angle=0, width=0.45\textwidth]{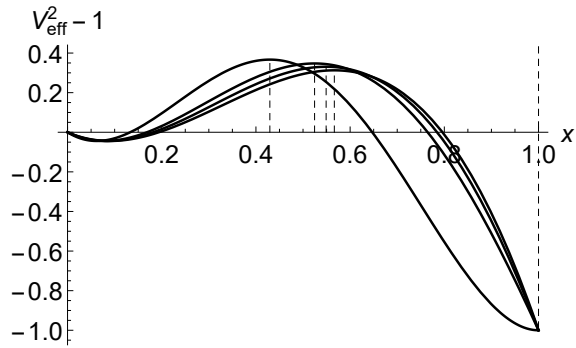}
    \vspace*{8pt}
\caption{The behavior of the effective potential with the different values of $\ell$. From left to right,  the dotted lines correspond to $\ell^{2}$ = $\frac{16}{27}$, $\frac{12}{27}$, $\frac{8}{27}$ and $\frac{3}{27}$, respectively. We choose $m = 1$, $b=5$. \label{v4}}
\end{figure}

\begin{figure}[H]
 \centering
    \includegraphics[angle=0, width=0.28\textwidth]{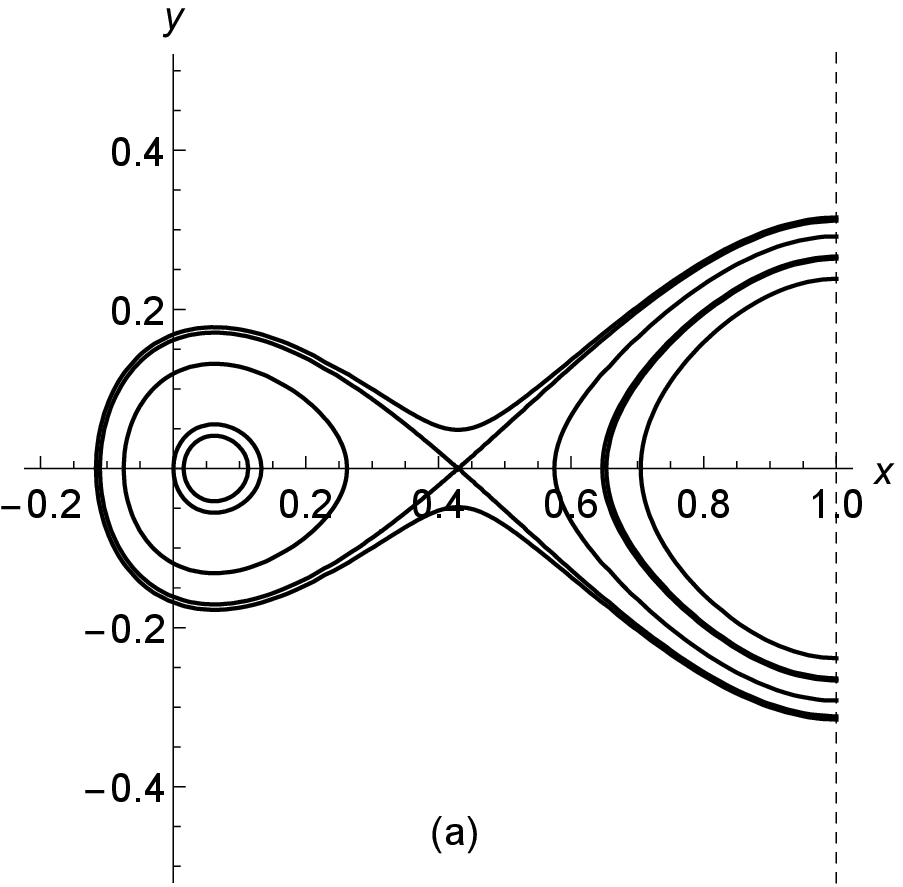}
    \includegraphics[angle=0, width=0.28\textwidth]{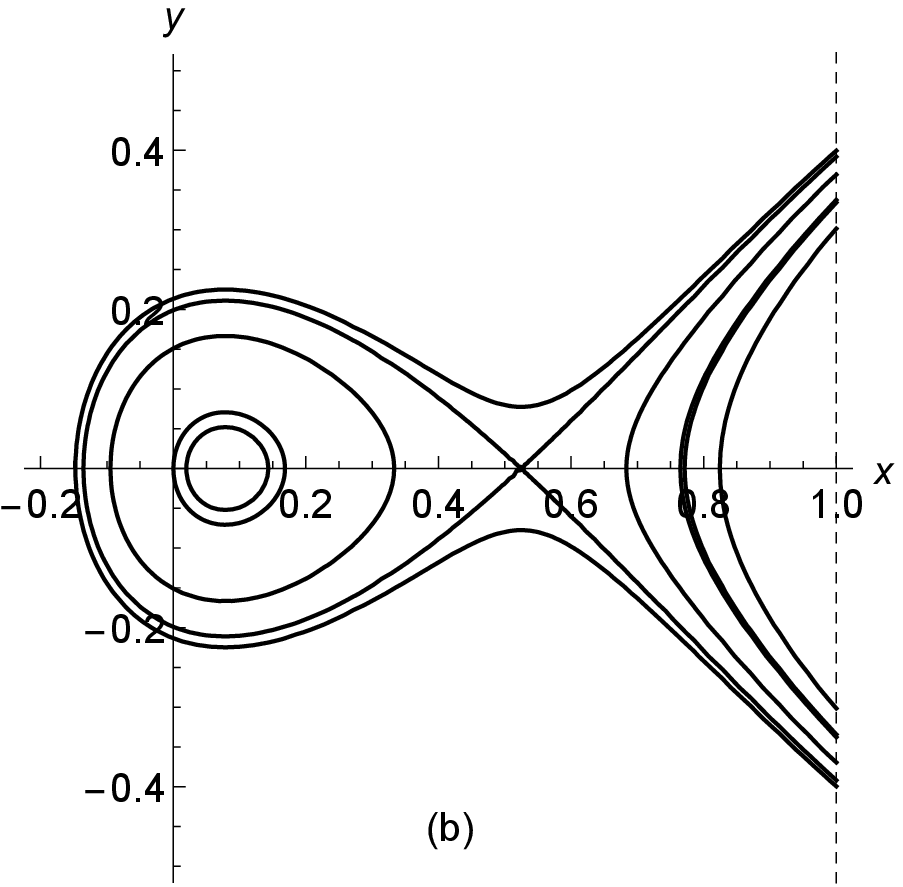}\\
     \ \includegraphics[angle=0, width=0.28\textwidth]{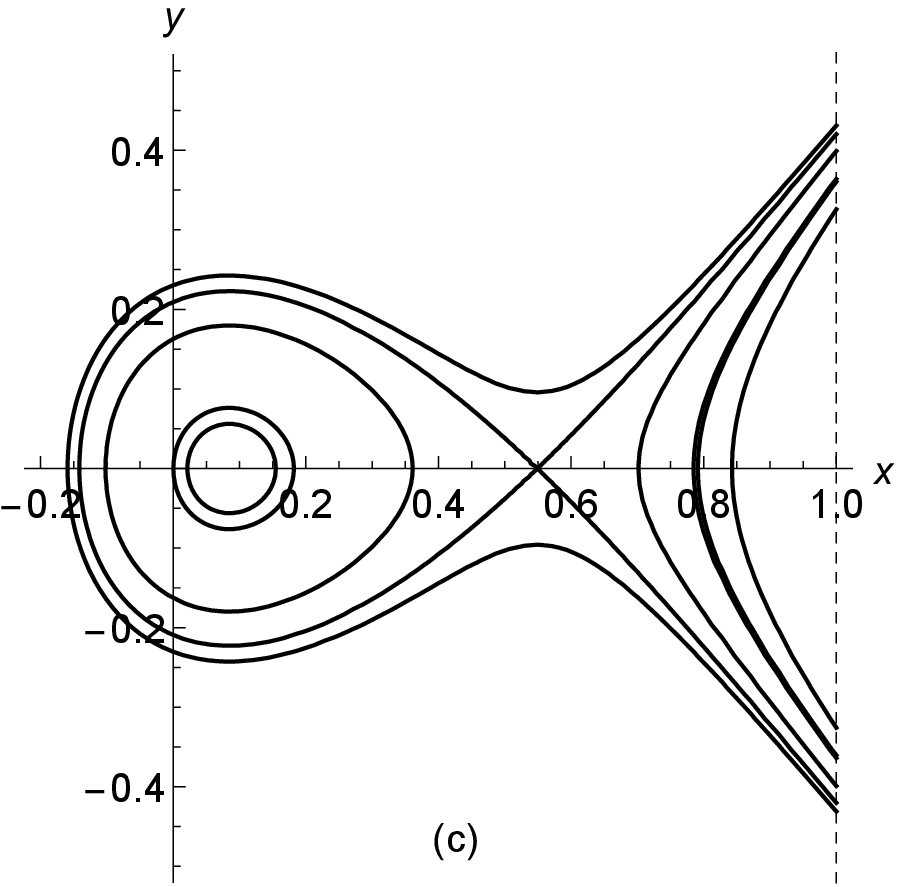}
        \includegraphics[angle=0, width=0.28\textwidth]{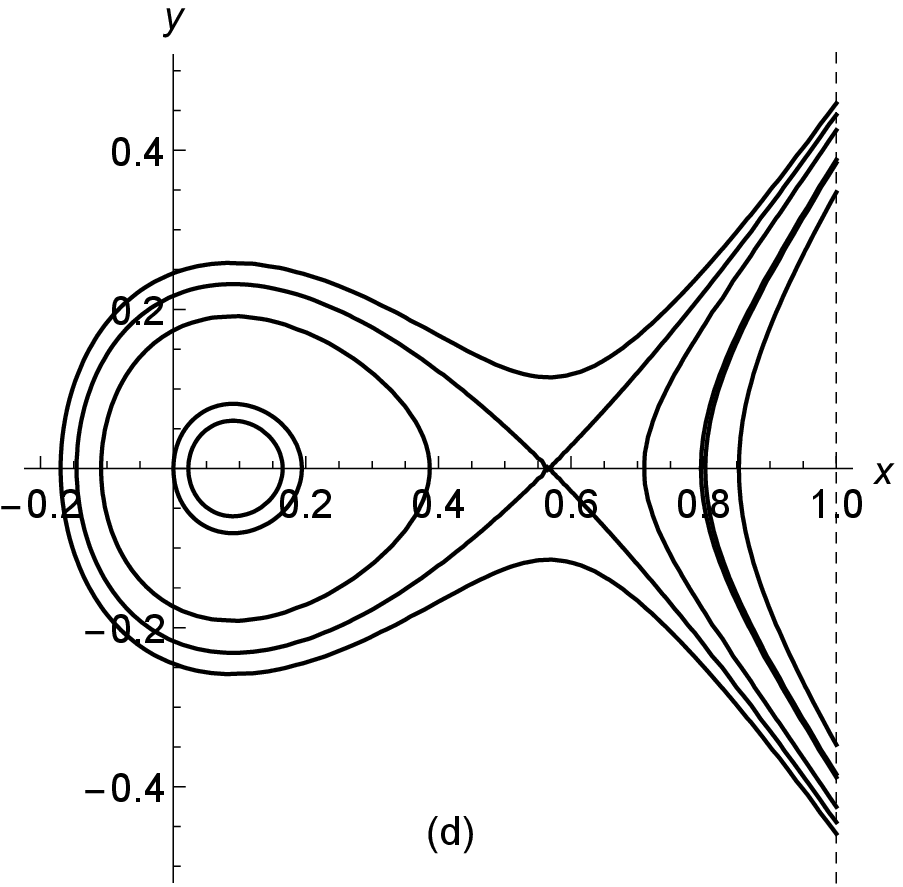}
        \vspace*{8pt}
\caption{The phase planes with different values of $\ell^{2}$. (a)$\ell^{2}$ = $\frac{16}{27}$, (b)$\ell^{2}$ =$\frac{12}{27}$, (c)$\ell^{2}$ =$\frac{8}{27}$ and (d)$\ell^{2}$ =$\frac{3}{27}$. We choose $m = 1$ , $b=5$. \label{v5} }
\end{figure}

In Fig. \ref{v4}, we find that two regions changed more clearly. In order to make a better understanding of this phenomenon, we plot the phase plane graph by using the phase plane method. Combining Figs. \ref{v4} and \ref{v5}, we make a brief analysis.

When the values of $\ell$ increases, (1) The range of stable orbits becomes larger, the test particles can move on a stable orbit which has a smaller radius; (2) The intersection point of X-axis and the dividing line of orbital stability moves left with $\ell$ increasing. The essence of this change is to reduce the innermost stable orbital radius. $r(a)_{min}=3.10391$, $r(b)_{min}=3.21825$, $r(c)_{min}=3.31732$ and $r(d)_{min}=3.42648$ correspond to energy values $(a) E^{2}-1$=0.366839, $(b) E^{2}-1$=0.347379, $(c) E^{2}-1$=0.330742, and $(d) E^{2}-1$=0.312809, respectively. We discover that in order to maintain the innermost stable orbit, $E$ needs to increase but $r$ ($x=\frac{r_{+}}{r}$) needs to decrease.

\section{Conclusion}
In this paper, we have analysed the orbital dynamics of test particles in regular Hayward black hole gravitational field. The orbital types are studied by analyzing the effective potential and using the phase plane method. We have discussed the influence of the energy $E$ and the angular momentum $b$ of the test particles on the orbital motion. It is found that the number of orbital types can be determined by the angular momentum $b$, the types of orbits are determined by the energy $E$. We have drawn a new and important conclusion that, the stability of orbital motion will vary when the values of $\ell^{2}$ changes from $\frac{3}{27}$ to $\frac{16}{27}$, through a global analysis of the effective potential graph and the phase plane graph. Our results show that: when $\ell$ increases, $E$ of the innermost stable orbit increases, and the radius $r$ decreases. When $E$, $b$ and $\ell$ are determined, the motion states of the test particles are also determined correspondingly. It is well known that gravitational lensing can be described by the study of orbital motion. By using the approximate geodesics as the basis for gravitational lensing, Kling et al.\cite{kling2000iterative} studied a single Schwarzschild lensing. Thus, we believe that the obtained results of this paper can provide some help for studying the gravitational lensing of the regular Hayward black hole.

{\bf Acknowledgments}

This work was supported in part by the National Natural Science Foundation of China (Grant No. 11565016), the Special Training Program for Distinguished Young Teachers of the Higher Education Institutions of Yunnan Province (Grant No. 1096837802).


\begin{thebibliography}{}
\bibitem{bardeen1968non} J. M. Bardeen: Proc. Int. Conf. GR5, Tbilisi 174, (1968).
\bibitem{virbhadra2007time} K. S. Virbhadra and C. R. Keeton: Phys. Rev. D 77, 124014 (2007).
\bibitem{frolov2008singularity} A. V. Frolov: Phys. Rev. Lett. 101, 061103 (2008).
\bibitem{khodadi2016more} M. Khodadi, K. Nozari and H. R. Sepangi: Gen. Relativ. Gravit. 48, 166 (2016).
\bibitem{hayward2006formation} S. A. Hayward: Phys. Rev. Lett. 96, 031103 (2006).
\bibitem{abbas2014geodesic} G. Abbas and U. Sabiullah: Astrophys. Space Sci. 352, 769-774 (2014).
 \bibitem{virbhadra2000schwarzschild} K. S. Virbhadra and G. F. R. Ellis: Phys. Rev. D 62, 084003 (2000).
\bibitem{cardoso2003quasinormal} V. Cardoso, R. Konoplya and J. P. S. Lemos: Phys. Rev. D 68, 044024 (2003).
 \bibitem{konoplya2005scalar} R. A. Konoplya and  E. Abdalla: Phys. Rev. D 71, 084015 (2005).
 \bibitem{Ibrar2016Marginally} I. Hussain and S. Ali: Eur. Phys. J. Plus 131, 275 (2016).
\bibitem{halilsoy2014thin} M. Halilsoy, A. Ovgun and S. H. Mazharimousavi: Eur. Phys. J. C 74, 2796 (2014).
\bibitem{bardeen1973four} J. M. Bardeen, B. Carter and S. W. Hawking: Commun. Math. Phys. 31, 161-170 (1973).
\bibitem{ayon1998regular} E. Ayon-Beato and A. Garcia: Phys. Rev. Lett. 80, 5056 (1998).
\bibitem{chiba2017anote} T. Chiba and M. Kimura: Prog. Theor. Exp. Phys. 2017, 043E01 (2017).
\bibitem{abdujabbarov2017gravitational} A. Abdujabbarov, B. Toshmatov, J. Schee, Z. Stuchl¨ªk and B. Ahmedov: Int. J. Mod. Phys. D 26, 1741011 (2017).
\bibitem{lin2013quasinormal} K. Lin, J. Li and S. Z. Yang: Int. J. Thero. Phys. 52, 3771-3778 (2013).
\bibitem{flachi2013quasinormal} A. Flachi and J. P. Lemos: Phys. Rev. D 87, 024034 (2013).
\bibitem{toshmatov2015quasinormal} B. Toshmatov, A. Abdujabbarov, Z. Stuchl\'{\i}k, and B. Ahmedov: Phys. Rev. D 91, 083008 (2015).
\bibitem{sharif2014stability} M. Sharif and S. Mumtaz: Adv. High Energy Phys. 2014, 639759 (2014).
\bibitem{cardoso2009geodesic} V. Cardoso, A. S. Miranda, E. Berti, H. Witek and V. T. Zanchin: Phys. Rev. D 79, 064016 (2009).
\bibitem{chen2009timelike} J. H. Chen and Y. J. Wang: Int. J. Mod. Phys. A 25, 1439-1448 (2010).
\bibitem{muller2011studying} T. M\"{u}ller and J. Frauendiener: Eur. J. Phys. 32, 747-759 (2011).
\bibitem{farrugia2017thermodynamic} C. Farrugia and J. Sultana: Gen. Relativ. Gravit. 49, 4 (2017).
\bibitem{azam2017geodesic} M. Azam, G. Abbas, S. Sumera and A. R. Nizami: Int. J. Geom. Method M. 14, 1750120 (2017).
\bibitem{azam2017geodesic2} M. Azam, G. Abbas and S. Sumera: Can. J. Phys. 95, 1062-1067 (2017).
\bibitem{amir2015rotating} M. Amir and S. G. Ghosh: JHEP 1507, 015 (2015).
\bibitem{amir2016collision} M. Amir, F. Ahmed and S. G. Ghosh: Eur. Phys. J. C. 76, 532 (2016).
\bibitem{hussain2015a} I. Hussain, M. Jamil and B. Majeed: Int. J. Theor. Phys. 54, 1567-1577 (2015).
\bibitem{dean1999phase} B. Dean: Am. J. Phys. 67, 78-86 (1999).
\bibitem{rosa2007stability} V. M. Rosa and P. S. Letelier: Phys. Lett. A 370, 99-103 (2007).
\bibitem{muslih2005hamiltonian} S. I. Muslih and  D. Baleanu: J. Math. Anal. Appl. 304, 599-606 (2005).
\bibitem{fernando2012schwarzschild} S. Fernando: Gen. Relativ. Gravit. 44, 1857-1879 (2012).
\bibitem{sheng2011time-like} S. Zhou, J. H. Chen and Y. J. Wang: Chinese Phys. B 20, 100401 (2011).
\bibitem{kling2000iterative} T. P. Kling, E. T. Newman and A. Perez: Phys. Rev. D 61, 104007 (2000).
\end{thebibliography}
\end{document}